\documentstyle[osa,manuscript,psfig]{revtex}

\begin{document}                                                          
\titlepage

\title{ 
%\hskip 7.0truecm {{{\normalsize FUB-HEP/96-11 (revised and extended 
%version)}}}\\
%	                \vspace {1.2truecm}
Probing the exchanged object(s) in diffractive scattering}
%\thanks{ Supported in part by Deutsche Forschungsgemeinschaft 
%(DFG: Me 470/7-1$
\author {C. Boros$^1$, Liang Zuo-tang$^{1,2}$ and Meng Ta-chung$^1$\\
 $^1$ Institut f\"ur theoretische Physik, FU Berlin, 
   Arnimallee 14, 14195 Berlin, Germany\\
 $^2$ Department of Physics, Shandong University, Jinan 250100, China }
 \maketitle    
                                                       
\begin{abstract}

The reaction mechanisms of the following proton 
diffractive scattering processes are
studied and compared with one another: 
$\gamma^* p\to Vp,\gamma^* p\to Xp, pp\to
(\Lambda K^+)p$ and $pA\to \Lambda K^+A$ 
(Here, $\gamma^*$ stands for the space-like photon 
in inelastic electron-proton scattering,
$V$ for the vector mesons $\rho,\omega,\phi,J/\psi$, and $A$ for a 
target nucleus.)
It is shown that, by taking parity,  and C--parity conservation 
into account, the existing data for these reactions 
 strongly suggest the following: 
A considerable part of the objects exchanged between projectile and
target in the above-mentioned  processes are virtual quark-antiquark pairs in
color-singlet states 
with  $J^{PC}=0^{-+}$, $I^G=0^+$, where 
$J$ stands for total angular momentum, 
$I$ for isospin, $P$, $C$ and $G$ stand for parity, C-parity and
G-parity respectively. Such quark-antiquark pairs are created by soft-gluon 
interactions. New experiments are
suggested; and predictions for such 
experiments are made. 

\end{abstract}

%\narrowtext
%\twocolumn
\newpage

Diffractive processes in lepton-nucleon 
scattering  at relatively large invariant
lepton-momentum transfer ($Q^2$) have received much attention 
recently [1-5]. Considerable efforts have been devoted to the problem 
whether, and if yes how,      
nucleon diffractive scattering in such  lepton-induced 
processes is related to that in  hadron-induced reactions. 
The questions in this problem-complex are usually 
discussed in the Regge-language: What are the similarities and/or  
the differences
between 
``the hard pomeron'' and ``the soft pomeron''?  
How does (do) the pomeron(s) 
interact  with the photon in 
photo-production and electro-production processes ? 
What are the properties of ``the pomeron structure function(s)'' ? etc..  
While the answers to these questions are still under debate [1-5], 
we think it may be useful also to discuss them in a different
language 
--- namely directly in terms of gluons and quarks in accordance with
QCD. 
To be more precise, stimulated by the ideas which have been 
discussed already for a
long time namely:
``pomeron-exchange dominates diffractive reactions'' [1-5] 
``pomeron can be interpreted as a system of gluons'' [6], 
we propose to study the mechanism(s) of exchange of gluon-systems 
in such processes directly.

We  consider inelastic lepton-nucleon scattering 
in the small $x_B$ ($< 10^{-2}$, say) region at a  given $Q^2$.  
It is in this kinematical region that
 large rapidity-gap events have been  observed [1-5] indicating 
the existence of diffraction. It is also in this region that
 structure function measurements  
 together with empirical analyses [7] show the following: the gluon-density 
for small $x_B$ is  much higher than that e.g. at 
$x_B\sim 0.1$ for the same $Q^2$. Hence, if such 
diffractive scattering processes are indeed due to ``exchange 
of systems of gluons'', these gluons are expected to be very ``soft'', in
the sense that even in a fast moving frame 
(e.g. the photon-proton c.m.s. frame) 
they  do not carry
 much  longitudinal momenta. 
Due to lack of sufficient energy and momentum, the chance for
such a ``soft'' gluon to fluctuate and  
create for a limited time-interval a quark-antiquark-pair is very small. 
But, due to the possibility of gluon-gluon interactions allowed by QCD 
in general, and the processes of virtual
quark-antiquark production by two or more gluons   
(as illustrated in Fig.1) in particular, we are led to consider, in diffractive
processes, the interaction between the virtual photon $\gamma^*$ and such virtual
quark-antiquark $(q\bar q)$ pairs.  
Because of the large rapidity gap, the $q\bar q$-pairs are expected to
be color singlets. What else do we know about such 
$q\bar q$-states ? What  
 quantum numbers can  such states have ?

In order to answer these two questions, it is useful to recall the
similarities and the differences of QCD and QED, in particular the dynamical 
constraints for $\gamma$-decay  of positronium and  
their  time-reversed processes, namely 
 fermion-pair production through photon-photon interactions.  
In analogy to QED [8], we see that parity-conservation and 
C-parity-conservation play important roles
 in describing the decay and the  formation  of  
$q\bar q$ pairs  in collision processes of soft-gluons. 
To be more precise, in analogy 
to positronium, a two-gluon state has C-parity $+1$, 
and therefore the only $q\bar q$ states that can decay into two 
gluons must have $L+S$ even, 
 where $L$ and $S$ stand for relative
angular momentum and total spin  
of the $q$ and the $\bar q$.  This implies that the 
lowest $C=+1$ state which can decay into two gluons is the ground-state
--- the singlet state characterized by $J=L+S=0$, parity $-$, and 
C-parity $+$. 
 This also implies: When two soft
gluons meet each other and form,  
for a finite time-interval, a color-singlet state, 
this  state is associated with the quantum 
numbers $J^{PC}=0^{-+}$, where $J$ stands for the total angular
momentum, P the parity and C the C-parity  
of the $q\bar q$ system. 
We also note that the quark $q$ may be u, d, s, c, ... and that isospin
conservation and G-parity conservation dictate:
The isospin ($I$) 
and the G-parity ($G$) of such $q\bar q$ systems should be $I^G=0^+$. 
%\vspace*{-1.5cm}
%\begin{figure}[h]
%\psfig{file=fig1.eps,width=10.cm}
%\caption{
%Fermion-pair ($f\bar f$) production  in gluon-gluon
%($gg$)-scattering.} 
%\end{figure}
It is expected that the 
$q\bar q$-pairs formed by two-gluon
interactions (see Fig. 1) play a dominating role, and this
expectation 
is based on the following simple considerations:
The interaction between soft-gluons is very complicated. 
Such gluons are expected to interact 
with one another according to QCD. But, because of the 
magnitude of the running coupling constant,  description of their interaction 
by using perturbative methods are no more reliable.  
All we can say in this connection is that, 
statistically, 
the chance for two soft-gluons to interact with each other and
form such a state is expected to be larger than that for three or more gluons to
come together. Besides, 
since direct gluon-gluon interactions 
 are allowed in QCD, $q\bar q$-creation
is also possible when, for example, 
 two gluons first directly interact with each other 
to become one gluon, and then produce the $q\bar q$-pair with a third gluon.

In order to answer the  question: 
``What do we expect to see experimentally, when such $\gamma^*$-$(q\bar q)$ 
interactions take place? '', 
we note that $q$ and $\bar q$ of
the above-mentioned $q\bar q$-system carry (opposite) electric charges.
Hence, for the incoming photon $\gamma^*$ which has a transverse dimension of
order $Q^{-2}$, $q$ and $\bar q$ may act 
--- depending on its transverse dimensions 
--- either (A) as one single object --- 
perhaps as an electrical dipole with 
extremely small dimensions 
(compared with the transverse dimension of $\gamma^*$ mentioned 
above), or (B) as a neutral 
system which consists of two opposite electric charges. 
We now discuss these two cases separately.

(A) What do we expect to see in 
 those cases in which  $Q^2$ of the 
virtual photon is small, such that its transverse 
dimension is large enough for 
the $q\bar q$-system  to act as one object and the transfer of momentum is not
so large to ``break up'' this system?
In order to answer this question, 
we  recall that the photon and the gluon have very
much the same properties as far as spin, parity 
and C-parity are concerned. 
In fact, 
taken together with the well-established conservation 
laws in electrodynamic and hadronic reactions [8], 
a space-like $q\bar q$-pair (with $J^{PC}=0^{-+}$) 
may absorb a virtual photon 
and become a time-like ($\gamma^*q\bar q$)-system 
with $J^{PC}=1^{--}$ and $I^G=0^-$ or $1^+$ 
(due to electromagnetic interaction which  conserves neither I nor G). 
That is to say, 
the final state of such a $\gamma^*$-$q\bar q$ 
reaction may be a vector-meson, 
provided that the $J^{PC}=0^{-+}$ $q\bar q$-system 
interacts with the
exchanged virtual photon $\gamma^*$ as an entire object! 
This means, we should see
vector-mesons in such lepton-nucleon processes in the small $x_B$ 
region, in particular: 
$$ e^-+p\to e^-+V+p$$
where $V$  stands for 
$\rho^0$, $\omega$, $\phi$, $J/\psi$. 

Furthermore,  we should see that 
the cross section for such processes 
is  proportional to the number density of 
such $q\bar q$-pairs associated with the proton. 
It  implies in particular that,    
at a given $Q^2$, the production rate of 
vector mesons are expected to increase with increasing $W^2$ 
(where $W$ is the total c.m. energy of the photon-proton system)
--- independent of the quantum numbers of the vector mesons. 
This is because, at fixed $Q^2$,
$$  W^2= Q^2(1/x_B-1)+M^2$$
(where M stands for the  proton-mass) 
increases with decreasing $x_B$. But, since  for smaller $x_B$ 
values, the gluon-density is higher [7], and thus 
the chance to produce such $J^{PC}=0^{-+}$ 
$q\bar q$-pairs is higher. This implies, the rate of production
of such $(\gamma^*q\bar q)$-systems (the vector mesons),
in particular the total cross-section 
$\sigma_T+\sigma_L$ for vector meson production is 
expected to become larger for increasing $W^2$. 
To compare this with the data quantitatively, 
we take the number-density of 
the colorless object (which consists of $q\bar q$) from 
 the data for the diffractive structure function 
$F^{D(3)}(\beta;x_B,Q^2)$ [1] and  integrate over $\beta$, 
(Here, $\beta$ stands for the fraction of the longitudinal 
momentum of the colorless object carried 
by the struck charged constituent). 
The obtained result shows that the number density of such 
colorless objects increases with decreasing $x_B$.  
The comparison between the 
data [2] and the calculated result is shown in Fig.2. As expected, the
agreement is reasonable. 
It should be mentioned in this connection that, since 
(according to the proposed picture)  
the $q\bar q$ system in the color-singlet $J^{PC} =0^{-+}$ state is
due to the interaction between two soft-gluons which exist only in the small
$x_B$ region, the $W$-dependence of the production rate of vector-mesons is
expected to apply only 
for sufficiently small $x_B (< 10^{-2}$, say) values. This
means, the vector meson production at large $x_B$-values, for example those of
NMC [3], are {\it not} ``the knocked-out color-singlet $(\gamma^*q\bar q)$ 
systems'' (which due to the interaction with $\gamma^*$ has $J^{\rm PC}=
1^{--}$).

Furthermore, for increasing $Q^2$, 
the production rate should decrease. 
This is  because the $q\bar q$-system is expected to break-up 
more easily when the transverse dimension $Q^{-2}$ of $\gamma^*$ becomes
smaller, so that the  geometrically more  point-like $\gamma^*$ can hit only 
either $q$ or $\bar q$. This, taken together with the fact 
that the transfer of momentum increases 
for increasing $Q^2$, show that 
we should see the following:
The $q\bar q$-system breaks-up   
much easier when $Q^2$ becomes large! 
As we can explicitly see in Fig.2 (note the $Q^2$-dependence 
of the cross-section for vector-meson production 
shown on the left-hand-side in the figure) and 
the corresponding multi-hadron production processes discussed 
in (B) below, 
also these characteristic features 
have indeed been observed [2,3]. 
%\begin{figure}
%\psfig{file=fig2.eps,width=8cm}
%\caption{Comparison between data [2,3] and the calculated results 
%(see text) for vector meson production.}
%\end{figure}

(B) Let us now examine in more detail the case in which $Q^2$ 
is so large, that (because of the reason mentioned above) it is 
more probable for $\gamma^*$ to  
hit only one of the charged objects
$q$ or $\bar q$. The struck object ($q$ or $\bar q$) can
absorb $\gamma^*$ and  ``fly away'' from 
the rest of the system (mainly $\bar q$ or $q$). Due to 
 color-forces,
the ``leading $q$ and $\bar q$ ''   is expected to behave 
like the $q\bar
q$-pair similar to that   
in high-energy electron-positron
scattering (see e.g. Ref.9) 
where a color-string is generated which breaks up into 
hadrons [10]. 
In order to see why these two kinds of processes are expected to be so 
similar to each other, 
 we note once again that, since 
a ($q\bar q$)-pair due to $gg$ is in the $J^{PC}=0^{-+}$ state, 
the ($\gamma^*$-$q\bar q$)-system (i.e.  
the system $q\bar q$ after one of its charged member has  absorbed 
the $\gamma^*$)
is in the $J^{PC}=1^{--}$ state. 
It means, the latter has the same quantum numbers of the 
(initial, intermediate and  final) state of the 
$e^-e^+\rightarrow \gamma^* \rightarrow q\bar q$ process.  
%\begin{figure}
%\psfig{file=fig3a.eps,width=8cm}
%\psfig{file=fig3b.eps,width=8cm}
%\caption{ (a) The $x_F$-distribution and (b) the 
%$\langle p_{\perp}^2\rangle$ vs $x_F$ plot for 
%the  hadrons  produced 
%in $\gamma^*$-$q\bar q$ collisions in the c.m.s. of 
%$\gamma^*$-$q\bar q$, for invariant mass $M_X$=10 GeV. 
%The data are taken from Ref.[1]; they are 
% given at fixed $M_X=10$ GeV 
%(in the kinematical range 2.5$< Q^2< $65 $GeV^2$, 0.01$<\beta <$0.9, 
%0.0001$<x_P<$0.05). The histograms are the 
%results of  a calculation  using JETSET [10], in which 
%the total c.m.s energy of the $e^-e^+$ system  
%is taken to be 10 GeV, and  the highest  
% value for the trust is taken to be 0.76, in  according 
%with the data given in [1]. } 
%\end{figure}
Having these arguments in mind, we reach the 
conclusion that,
 in such cases, we should see the following:  
The produced hadrons are   distributed 
along the scattering axis symmetric  
 with respect to the center of mass of the  $\gamma^*$-$(q\bar q)$ system. 
 Only the hadrons on the end of the color-string can have 
large momenta (both the longitudinal component 
 $p^*_{||}$ and the transversal components $p_{\perp}^*$  
with respect to the collision axis).
 In other words, if we follow H1-Coll. [1] and 
define $x_F=2p^*_{||}/M_X$ where $M_X$ is
the invariant mass of the system of produced hadrons, we should see that
most of the hadrons are concentrated near $x_F=0$ and have very small
$p_{\perp}^{*2}$. In order to compare the proposed 
picture with the data [1] in a quantitative manner, we use the Lund
model as implemented in JETSET [10] to calculate the $x_F$-distribution,   
the $\langle p_{\perp}^2\rangle$ vs. $x_F$ plot and 
the energy flow as a function of pseudo-rapidity in the center 
of mass system of the photon and the exchanged object.  
The obtained results are shown
in Fig. 3a and 3b. and Fig.4. 
It should be  mentioned that, while the data in Figs. 3(a) and 3(b) are
given [1] at fixed $M_X$ ($M_X=10$ GeV), those in Fig.4 are given 
[1] for three different ranges of $M_X$. 
Hence, taken together with the 
the kinematical relationship 
$M_X^2\approx Q^2 (x_P/x_B-1)$  which shows that 
$Q^2$ and $M_X^2$ are directly proportional to each 
other for fixed values of $x_P$ and $x_B$,  
we can at least use Fig.4 to check the $Q^2$-dependence 
of the proposed picture for fixed $x_B$ and $x_P$. 
Having in mind that the transverse dimension of 
$\gamma^*$ becomes smaller 
and the transfer of momentum to the struck quark or the antiquark increases
with increasing $Q^2$ and thus with increasing $M_X^2$,  
we expect to see that the multi-hadron events should be
more jet-like for increasing $M_X^2$. 
As we can see in Fig.4, this expectation is indeed in agreement with
the data [1]. It would be interesting to see whether this trend will continue
for even larger $M_X$-values in future experiments.
%\begin{figure}
%\psfig{file=fig4.eps,width=10.cm}
%\protect
%\caption{The energy flow is plotted as function of the pseudo-rapidity 
%\protect{$\eta^*$} with respect to the \protect{$\gamma$}-
%\protect{$q\bar q$} axis in the 
%center of mass system of \protect{$\gamma$}-\protect{$q \bar q$}. 
%The  curves are the calculated results for quark anti-quark 
%fragmentation using  JETSET 
%[10]. 
%Here, integration has been performed in the 
%corresponding \protect{$M_X$} regions where 
%the energy-flow at each given \protect{$M_X$} is 
%weighted  by the factor \protect{$1/M_X^2$} 
%(which is  known [18] to be 
%the characteristic  \protect{$M_X$}-distribution in  diffractive 
%scattering).  
%Note that, according to 
%the data [1], the mean thrust should 
%be 0.7 for the first, 0.75 for the second 
%and 0.82 for the third \protect{$M_X$}-region.  
%This piece of experimental fact has also been taken into account in the 
%present calculation. The data are from [1].}
%\end{figure}

We next turn 
our attention to diffractive hadron-hadron collisions, and ask: Do
 we also see experimental indications for the exchange of   
spin-zero virtual $q\bar q$-systems due to interactions between
soft-gluons in such collision processes?
One way of probing the spin of the exchanged $q\bar q$-pair in diffractive
scattering is to look at proton diffractive dissociation processes in which the
final states are simple.
In this sense, the reactions $pp\to (\Lambda K^+)p$ and 
$pA\to (\Lambda K^+)A$  are ideal. 

Such experiments  
have already been performed [11,12]. Very striking polarization effects 
have been observed 
in the former [11] and similar measurements are underway in the latter
[12]:  
The produced $\Lambda$ observed 
in [11] is polarized (with respect to the axis 
$\vec p_B \times\vec p_\Lambda$ 
where  $\vec p_B$ and $\vec p_\Lambda$ are the 3-momenta of the proton
beam 
and the outgoing $\Lambda$ respectively) although neither the projectile nor the
target is polarized. The polarization of the $\Lambda$ is very large
$(62\% \pm 4\%)$ -- much larger than those observed in inclusive reactions
$pp\to \Lambda X$ [13]. This piece of experimental fact, together with 
 strangeness conservation in hadronic
processes,  makes  the study of
this process even more interesting.

We recall
that the striking features especially the flavor-dependence, the
projectile-dependence of asymmetries in single-spin hadron-hadron
collisions 
can be   described in terms of a relativistic
quark-model\cite{[14]}  in which valence quarks are treated
as Dirac particles confined in a spatially extended
hadron.    
It is  shown
that such valence quarks perform orbital motion,
the spin states of the valence quarks
with respect to the polarization
axis of a polarized hadron  are determined by the wave function of the
hadron, and thus
a meson directly formed by a valence quark
near the front surface of the polarized hadron
(with respect to the target hadron which provides a suitable
antiseaquark)
acquires an extra amount of transverse momentum due to the
orbital motion of the valence quark.  
{\it This
implies the existence of a close relationship
between the polarization of the
valence quarks
of the transversely polarized
projectile hadron and
the transverse motion  of the
directly formed mesons in the projectile fragmentation
region.
In other words, in this picture, knowing the probability of observing
a meson on one side (left or right) of the collision axis, the chance for the
valence quarks of the projectile to be polarized in a given direction
perpendicular
to scattering plane is determined.} 
Having this picture in mind, we see that the $J^{PC}$ of the $s\bar
s$-pair plays a decisive role: 
As graphically demonstrated 
in Fig.5, the $\Lambda$ in 
$pp\rightarrow (\Lambda K^+)p$ is expected to be negatively 
polarized with respect to the production plane.  
To be more precise, the 
produced $K^+$ in this kinematical region is the fusion 
product of a u-valence quark and a strange antiquark $\bar s$ form the 
sea. 
Because of the correlation between the polarization and the
transverse momentum of the valence quarks, as explained 
above (and in more detail in Ref.14), the 
 valence quark $u_v$ has a
larger probability to be polarized downwards (than upwards). Since
$K^+$ is a spin-0-particle, the $\bar s_s$ quark 
originating from the above-mentioned $J^{\rm PC}=0^{+-}$ system
is polarized
upwards.  The $\Lambda$ which 
goes left (because of momentum conservation) is directly formed by the
remaining
$ud$-diquark of the projectile proton and the
downwards polarized s-quark $s_s$.
Since the $ud$-diquark is a spin-0-system, the
 polarization of
 $\Lambda$  is determined by the polarization of  $s$-quark.
 It is opposite to $\vec{n}\equiv \vec{p_B}\times \vec{p_{\Lambda}}/
 |\vec{p_B}\times \vec{p_{\Lambda}}|$ and hence negative. 
 Needless to say that, the conclusion we reached does not depend 
on the fact that we have chosen to consider the $K^+$'s  moving
to the right
with respect to the collision axis 
(as shown in this figure). 
By considering those moving to the left,
by 
replacing ``downwards" by ``upwards" and vice versa in the discussions
correspondingly, and by keeping the definition of $P(\Lambda)$ in mind, we
obtain the same result. 
Here, we  see among other things 
that, in the framework of the relativistic 
quark model [14], 
$P_{\Lambda}$ for the diffractive process 
$pp\rightarrow (\Lambda K^+)p$ {\it could never} be larger than that for
the inclusive 
$pp\rightarrow \Lambda X$, if the $s\bar s$-pair 
{\it were not} spin-zero objects!  
%\begin{figure}
%\psfig{file=fig5a.eps,width=10cm}
%\psfig{file=fig5b.eps,width=9cm}
%\caption{
%Proton diffractive dissociation $pp\rightarrow (\Lambda K^+)p$, where
%the produced
%$K^+$ is observed on the right-hand-side of the
%$p+p$  collision axis. 
%(The conclusion that the $\Lambda$ is negatively 
%polarized with respect to the production 
%plane is of course independent 
%of the fact, whether the $K^+$ is observed 
%on the right-hand-side. See text for more details.)  }
%\end{figure} 

Two further remarks should be made in this connection: 

First, in a recent paper [15], J. Felix 
et al published their data on 
$p p\rightarrow p\Lambda K^+ \pi^+\pi^-\pi^+\pi^-$ in which 
the following characteristic features have been observed. 
(a) $P_{\Lambda}$ is significantly smaller ($\sim 15\%$) than 
that observed in [11]  
(b) The polarization of $\Lambda$ is independent of the fact 
whether $K^+$ is detected in the same or in the 
opposite hemisphere as $\Lambda$. 
Both (a) and (b) are consistent 
with the proposed picture. This is because, in contrast to the 
$pp\rightarrow K^+ \Lambda p$ experiment [11] which is performed 
at $p_{inc}\approx 2000$ GeV/c, the experiment 
$p p\rightarrow p\Lambda K^+ \pi^+\pi^-\pi^+\pi^-$ is done at 
$p_{inc}\approx 27.5$ GeV/c. 
Here, we note in particular that, although the total center-of-mass-energy
is so low, the number of particles in the final state of this 
reaction is rather  high. 
Hence the chance for  this process to undergo 
a single diffractive process is relatively small.  
Having in mind that the exchange of a $gg$-system with the above 
mentioned quantum numbers dominates  only in single particle 
diffractive processes it is expected that the $\Lambda$-polarization 
in this process [15] should be significantly 
smaller than that observed in [11]. 

In connection with the experimental fact mentioned in 
(b) we  discuss the 
contributions from single particle  diffractive processes to 
$p p\rightarrow p\Lambda K^+ \pi^+\pi^-\pi^+\pi^-$ 
and show that the $\Lambda$ polarization is expected to be 
independent of the rapidity-distribution 
of $K^+$ even if the observed final state hadrons are contributions 
form single diffractive processes. 
To be more precise, in 
Fig.6 we show the final states due to the exchange 
a $J^{PC}=0^{-+}$, $I^G=0^+$ gg-state including a number 
of light quarks and antiquarks $u\bar u$ or $d\bar d$. 
Obviously, in such cases, 
 different possibilities of 
combining the $u\bar u$, $d\bar d$ and $s\bar s$ pairs to form the 
observed final states have to be taken into account. 
Here, the dark sites in these figures stand for ``black boxes''. 
We do not know, and do not care of the details
about the reaction 
mechanisms inside such black boxes,  
because they 
 do not play a role in the proposed 
picture. 
In fact, it is only important that the black box 
describes a spin-zero state. 
As illustrative examples, we show in Fig.6 two different 
possibilities to produce the final state 
hadrons $\Lambda K^+ \pi^+\pi^-\pi^+\pi^-$ in 
$p$-$p$-collision by the exchange of 
a colorless spin-zero $gg$-system. 
In (i) both   $\Lambda$ and $K^+$ are produced 
by  valence-quarks of the projectile. 
The probability is large that both of them appear 
in the fragmentation region of the projectile i.e. in the same 
hemisphere.    
Here $K^+$ compensates both the 
transverse momentum and the spin of the produced 
$\Lambda$ and give raise to the $\Lambda$-polarization. 
In (ii) one of the produced pions  and the $\Lambda$ are produced 
by the valence quarks of projectile. 
Since, here, the kaon is produced by sea-quarks 
there is no kinematical constraints which could force it 
to be in the same hemisphere as the $\Lambda$. In fact 
since one of the 
pions takes over the role played by $K^+$ in case (i) 
in compensating the transverse momentum  and 
the spin of the $\Lambda$, 
the $K^+$ in this case is free --- 
in the sense that it can be either in the same or 
in the opposite hemisphere relative to $\Lambda$!  
This means, 
 as far as the 
$\Lambda$-polarization is concerned  
the spatial correlation between $\Lambda$ and $K^+$  
(in particular whether they are in the same 
or opposite hemisphere) does not play a role when other 
spin-zero particles are produced. As we see in Ref. [15], this is 
  in accordance 
with the experimental findings. 
%\begin{figure}
%\psfig{file=fig6.eps,width=10.5cm}
%\caption{Two possibilities to produce the 
%final state:
%(i) Both  $\Lambda$ and $K^+$ are produced 
%by the valence-quarks of the projectile.   
%Here $K^+$ compensates both the 
%transverse momentum and the spin of
%$\Lambda$.  
%(ii) One of the produced pions takes over the role of $K^+$ 
%in compensating the transverse momentum  and 
%the spin of the $\Lambda$. }
%\end{figure} 

Second, further
experimental studies of polarization effects {\it in diffractive
dissociation processes,} 
for example in\\ $p(\uparrow )p\rightarrow K^+\Lambda (\uparrow ) p$ 
in which the ``beam-$\Lambda$-parameter'' $D_{NN}$ 
 is measured, will be  helpful especially 
in probing the spin of the
exchanged $s\bar s$-pair in diffractive scattering processes. 
If all exchanged colorless objects have spin 
zero, we expect $D_{NN}=1$. Hence the observed 
$D_{NN}$ in such diffractive processes is --- in the proposed 
picture --- a direct measure of the fraction of spin-zero 
components in the set of all exchanged 
colorless objects. 

Last but not least, the following should also be mentioned. 
While the $\Lambda$-particles in the fragmentation regions in 
proton-proton collisions --- especially in proton 
diffractive dissociation processes --- are predominantly 
created through quark-diquark fusion, 
the production mechanisms of $\Lambda$ in diffractive deep-inelastic 
electron-proton scattering is very much different: 
In the latter case, the (large $Q^2$) virtual photon $\gamma^*$ 
hits either s or $\bar s$   
of the above-mentioned color-singlet 
spin-zero $s\bar s$-system 
(formed by collisions of the two soft gluons) 
and produce a color-string led by  $s$ and $\bar s$. 
Because of helicity-conservation in high-energy 
photon-quark scattering 
 the helicity of the struck $s$ 
(or $\bar s$)  has to flip. 
It implies, after the $\gamma^*$-($s\bar s$) interaction,  
the $s$ and $\bar s$ should have different 
helicities in the center-of-mass frame of 
the $\gamma^*$-$(s\bar s)$ system. 
%%%%%%%%%%% 
Having in mind, that the 
 leading s and $\bar s$ can give rise to 
$\Lambda$ and $\bar\Lambda$ respectively,  
and that the polarization of $\Lambda$ ($\bar\Lambda$) is completely
determined by that of the $s$ ($\bar s$), the $\Lambda$ and 
$\bar\Lambda$ are expected to have opposite helicities. 
This means, by measuring the helicities of the $\Lambda$ and $\bar\Lambda$ 
appearing in opposite hemispheres on an event by event basis,  
we can experimentally test whether the struck $s\bar s$-system is a spin-zero state. 
%\begin{figure}
%\psfig{file=fig7.eps,width=9cm}
%\caption{The probability $P(h_\Lambda =-h_{\bar\Lambda })$ for the 
%observed $\Lambda$ in one hemisphere and the observed $\bar\Lambda$ 
%in the other hemisphere in a given event 
%are correlated, in the sense that they  have opposite helicities, is
%plotted 
%as a function of $(x_F)_{cut}$. Here, $(x_F)_{cut}$ 
% is the minimum of the magnitude of the $x_F$ of 
%$\Lambda$ and that of $\Lambda$.} 
%\end{figure} 

In order to estimate the probability for the $\Lambda$-$\bar\Lambda$ 
to have opposite helicities {\it quantitatively}, 
we use the method introduced be Gustafson and 
H\"akkinen [16]. 
As we know, by using the above-mentioned method, these authors have 
successfully predicted [16] the longitudinal
polarization measured in the recent ALEPH-experiment[17]. 
Following this method, we  consider all possible 
processes which can contribute to the production of a 
 $\Lambda$-$\bar\Lambda$ pair. They are the following: 

(a) Both the $\Lambda$ and $\bar \Lambda$ are directly formed 
by  the {\it initial} $s$ and $\bar s$ quarks 
respectively. The $\Lambda$ 
($\bar\Lambda$) has the helicity of the original $s$($\bar s$)-quark. 
In this case the above-mentioned  helicity correlation is maximal. 
The $\Lambda$ and the $\bar\Lambda$ have opposite helicities 
provided that the $s$-$\bar s$ pair is a spin-zero object. 

(b) The $\Lambda$ ($\bar\Lambda$) is produced {\it during the  
fragmentation processes of $s$ and $\bar s$} 
in which the $\Lambda$ ($\bar\Lambda$) 
does not contain the initial $s$ ($\bar s$) quark. 
It is obvious that, in such processes,  
the helicities of the $\Lambda$ and $\bar\Lambda$ 
 are {\it not} correlated. 

(c) Since such colorless clusters can also be $u$-$\bar u$  or 
$d$-$\bar d$ systems so that the produced lambda hyperons can also be 
the fragmentation products of such light quarks.  
Having in mind the polarization of $\Lambda$ ($\bar\Lambda$) is 
completely determined by that of the $s$ ($\bar s$) quarks, 
it is clear that these
events {\it do not} contribute to the $\Lambda$-$\bar\Lambda$ 
helicity correlations. 

(d) Produced $\Lambda$-$(\bar\Lambda$)  hyperons which contain  
initial $s$ ($\bar s$) quarks but are decay-products of 
heavier hyperon resonances. The helicities of 
these hyperon anti-hyperon pairs are in general also correlated if they 
originate from initial spin-zero $s$-$\bar s$ pairs. 
They decay into $\Lambda$ - $\bar\Lambda$ pairs which will ``remember'' 
part of the helicity 
correlation of the initial $s$-$\bar s$. 

In order to treat
 the contribution of those events associated with resonance-decay
quantitatively,  
we calculate the probabilities for the initial 
helicity correlation of the $s$-$\bar s$ pair to be 
transfered through the intermediate resonances to the final $\Lambda$- 
$\bar\Lambda$ pair.  
Let us first discuss   the production of $\Sigma^0$ which decays 
electromagnetically into $\Lambda+\gamma$. 
Here, we  recall that according to the method used in Ref.[16] 
when dealing with such heavy strange particles 
the non-relativistic quark model can be used to calculate the
above-mentioned 
probabilities. In fact, it is easy to see that 
the probabilities 
for  the produced $\Sigma$ to have the same  or opposite helicity as the
original $s$-quark are $P(\Sigma^0,s,+)=1/3$ and
$P(\Sigma^0,s,-)=2/3$ respectively. 
For the decay-process  
$\Sigma^0\rightarrow \Lambda+\gamma$ the probabilities 
that the $\Lambda$ should have the same or opposite helicity as the 
$\Sigma^0$  are $P(\Lambda,\Sigma^0,+)=1/3$ and 
$P(\Lambda,\Sigma^0,-)=2/3$ respectively. 
This means the probabilities for the $\Lambda$ originating  from the 
original $s$-quark through $\Sigma^0$ decay to have 
the same  or opposite helicity as the s-quark are given by: 
\begin{eqnarray}
 P(\Lambda^{\Sigma^0},s,+)&=& P(\Lambda,\Sigma^0,+) P(\Sigma^0,s,+), 
 \nonumber\\
 &+& P(\Lambda,\Sigma^0,-) P(\Sigma^0,s,-)=5/9 \\
 P(\Lambda^{\Sigma^0},s,-)&=& P(\Lambda,\Sigma^0,+) P(\Sigma^0,s,-) 
 \nonumber\\ &+& P(\Lambda,\Sigma^0,-) P(\Sigma^0,s,+)=4/9 
\end{eqnarray}
respectively. 
The corresponding probabilities  for $\bar\Lambda$ originating from the 
$\bar\Sigma^0$ decay are the same. 
The probabilities for that the $\Lambda$-$\bar\Lambda$ pairs 
which come from $\Sigma^0$-$\bar\Sigma^0$ decays to have the opposite 
or the same helicities if the $s$-$\bar s$ have opposite helicities 
are given by: 
\begin{eqnarray}
P(h_\Lambda=-h_{\bar\Lambda})&=&
P(\Lambda^{\Sigma^0},s,+) P(\bar\Lambda^{\bar \Sigma^0},\bar s,+)\nonumber\\ 
&+& P(\Lambda^{\Sigma^0},s,-)P(\bar\Lambda^{\bar \Sigma^0},\bar s,-)
=\frac{41}{81}\\ 
P(h_\Lambda=h_{\bar\Lambda})&=&
P(\Lambda^{\Sigma^0},s,+) P(\bar\Lambda^{\bar \Sigma^0},\bar s,-)\nonumber\\ 
&+& P(\Lambda^{\Sigma^0},s,-)P(\bar\Lambda^{\bar \Sigma^0},\bar s,+)
=\frac{40}{81}
\end{eqnarray} 
Next we note that, according to the 
analyses carried out by Gustafson and 
H\"akkinen[16],  the hyperon decays $\Sigma^0\rightarrow\Lambda\gamma$, 
$\Sigma(1385)\rightarrow\Lambda\pi$, 
$\Xi\rightarrow\Lambda\pi$ and 
$\Xi(1530)\rightarrow\Xi\pi\rightarrow\Lambda\pi\pi$ should be included 
because their contributions are relatively large and 
other possible decays should be neglected 
because their contributions are very small [16].
In Table 1. we list the corresponding probabilities 
for the other decays. For the anti-hyperons we get the same numerical 
results.  
In order to calculate the relative probabilities for 
observing $\Lambda$ and $\bar\Lambda$ which  
originate from (a), (b), (c) and (d) we again use the 
Lund-model as implemented in JETSET for the 
fragmentation of the $s-\bar s$-system. 
Precisely speaking, we calculate the probability for  
$\Lambda$ and $\bar\Lambda$ 
to have opposite helicities as a function of $(x_F)_{cut}$ which is 
the minimum of the $x_F$ of $\Lambda$ and that of $\bar\Lambda$. 
The obtained result is shown in Fig.7. Here, we 
  see that for large 
$(x_F)_{cut}$ the probability for the $\Lambda$ and the $\bar\Lambda$ 
to have different helicities is large. 
In the small $(x_F)_{cut}$-region in which 
 the production processes (b) and (c) 
dominate, there is practically {\it no} correlation between the 
helicities of the hyperons. In other words, the probability 
turns out to be 
$p(h_{\Lambda}=-h_{\bar\Lambda})\approx 0.5$, as it should be.
Note that, no correlation between the helicity $h_\Lambda$ of 
$\Lambda$ and the helicity $h_{\bar\Lambda}$ of $\bar\Lambda$ means
nothing else but 
that $h_\Lambda=h_{\bar\Lambda}$ and 
$h_\Lambda=-h_{\bar\Lambda}$ are equally probable. 
In other words, the probability $p(h_\Lambda =-h_{\bar\Lambda}$) for
$\Lambda$ and $\bar\Lambda$ to have opposite helicities should be
$50\%$.

In conclusion, the present paper points out that
 there exist a large number of theoretical 
arguments and experimental evidences which show the following: 
A considerable part of the ``exchanged colorless object(s) in the
diffractive scattering 
processes, in particular in 
$\gamma^*p\rightarrow Vp$, $\gamma^*p\rightarrow Xp$, 
$pp\rightarrow (\Lambda K^+)p$ and 
$pA\rightarrow (\Lambda K^+)A$, where $\gamma^*$ stand for virtual photon, 
$p$  for proton, and $V$ for vector meson, have 
the following quantum numbers: 
$J^{PC}=0^{-+}$, $I^G=0^+$ where 
$J$ stand for for the total angular momentum, 
$I$ for isospin, $P$, $C$ and $G$ stand for parity, C-parity and
G-parity respectively. Such color-singlet objects which exist for a
finite time-interval are formed by soft gluons the density of which
are expected to be very high in the small $x_B$ region of
lepton-nucleon scattering processes.
\vspace{1.cm} 

Acknowledgments:
 
The authors thank K. Heller, R. Rittel and Y. Wu for 
helpful discussions;
G.Gustafson and J.H\"akkinen for correspondence. 
This work is supported in part by 
Deutsche Forschungsgemeinschaft 
(DFG:Me 470/7-2) and FNK of FU-Berlin (FPS-Cluster).

\newpage

\begin{table*}
\caption{The probabilities for $\Lambda$ originating from 
different hyperon decays to have the same(+) [opposite(-)] 
polarization as the initial s-quark. See text for more details.} 
\vspace*{0.5cm}
\begin{tabular}{||l||c|c|c|c|c|c||}
\hline \hline 
\rule[-0.3cm]{0.0cm}{0.8 cm}
\phantom{xxx} & $P(H,s,+)$ & $P(H,s,-)$ & $P(\Lambda,H,+)$ & 
$P(\Lambda,H,-)$ & $P(\Lambda^H,s,+)$ & $P(\Lambda^H,s,+)$\\
\hline \hline 
\rule[-0.3cm]{0.0cm}{0.8 cm}
$\Sigma^0\rightarrow\gamma\Lambda$ & 1/3 
& 2/3 & 1/3 & 2/3 & 5/9 & 4/9 \\
\hline
\rule[-0.3cm]{0.0cm}{0.8 cm}
$\Sigma (1385)\rightarrow\Lambda\pi$ & 5/9
& 4/9 & 1 & 0  & 5/9 & 4/9 \\
\hline
\rule[-0.3cm]{0.0cm}{0.8 cm}
$\Xi\rightarrow\Lambda\pi$ & 5/6
& 1/6 & 9/10 & 1/10  & 23/30 & 7/30 \\
\hline
\rule[-0.3cm]{0.0cm}{0.8 cm}
$\Xi(1530)\rightarrow\Xi\pi\rightarrow\Lambda\pi\pi$ & 5/9
& 4/9 & 9/10 & 1/10  & 49/90 & 41/90 \\
\hline\hline
\end{tabular}
\end{table*}

\begin{figure}
\caption{Fermion-pair ($f\bar f$) production  in gluon-gluon
($gg$)-scattering.} 
\end{figure}

\begin{figure}
\caption{Comparison between data [2,3] and the calculated results 
(see text) for vector meson production.}
\end{figure}

\begin{figure}
\caption{ (a) The $x_F$-distribution and (b) the 
$\langle p_{\perp}^2\rangle$ vs $x_F$ plot for 
the  hadrons  produced 
in $\gamma^*$-$q\bar q$ collisions in the c.m.s. of 
$\gamma^*$-$q\bar q$, for invariant mass $M_X$=10 GeV. 
The data are taken from Ref.[1]; they are 
 given at fixed $M_X=10$ GeV 
(in the kinematical range 2.5$< Q^2< $65 $GeV^2$, 0.01$<\beta <$0.9, 
0.0001$<x_P<$0.05). The histograms are the 
results of  a calculation  using JETSET [10], in which 
the total c.m.s energy of the $e^-e^+$ system  
is taken to be 10 GeV, and  the highest  
 value for the trust is taken to be 0.76, in  according 
with the data given in [1]. } 
\end{figure}

\begin{figure}
\caption{The energy flow is plotted as function of the pseudo-rapidity 
\protect{$\eta^*$} with respect to the \protect{$\gamma$}-
\protect{$q\bar q$} axis in the 
center of mass system of \protect{$\gamma$}-\protect{$q \bar q$}. 
The  curves are the calculated results for quark anti-quark 
fragmentation using  JETSET 
[10]. 
Here, integration has been performed in the 
corresponding \protect{$M_X$} regions when
the energy-flow at each given \protect{$M_X$} is 
weighted  by the factor \protect{$1/M_X^2$} 
(which is  known [18] to be 
the characteristic  \protect{$M_X$}-distribution in  diffractive 
scattering).  
Note that, according to 
the data [1], the mean thrust should 
be 0.7 for the first, 0.75 for the second 
and 0.82 for the third \protect{$M_X$}-region.  
This piece of experimental fact has also been taken into account in the 
present calculation. The data are from [1].}
\end{figure}

\begin{figure}
\caption{
Proton diffractive dissociation $pp\rightarrow (\Lambda K^+)p$, where
the produced
$K^+$ is observed on the right-hand-side of the
$p+p$  collision axis. 
(The conclusion that the $\Lambda$ is negatively 
polarized with respect to the production 
plane is of course independent 
of the fact, whether the $K^+$ is observed 
on the right-hand-side. See text for more details.)  }
\end{figure}

\begin{figure}
\caption{Two possibilities to produce the 
final state:
(i) Both  $\Lambda$ and $K^+$ are produced 
by the valence-quarks of the projectile.   
Here $K^+$ compensates both the 
transverse momentum and the spin of
$\Lambda$.  
(ii) One of the produced pions takes over the role of $K^+$ 
in compensating the transverse momentum  and 
the spin of the $\Lambda$. }
\end{figure}

\begin{figure}
\caption{The probability $P(h_\Lambda =-h_{\bar\Lambda })$ for the 
observed $\Lambda$ in one hemisphere and the observed $\bar\Lambda$ 
in the other hemisphere in a given event 
are correlated, in the sense that they  have opposite helicities, is
plotted 
as a function of $(x_F)_{cut}$. Here, $(x_F)_{cut}$ 
 is the minimum of the magnitude of the $x_F$ of 
$\Lambda$ and that of $\Lambda$.} 
\end{figure}


\begin {thebibliography}{99}

\bibitem{[1]}
See e.g. J.P. Phillips, Talk given at the 28th 
Int. Conf. 
on High-Energy Physics, July 1996, Warsaw, Poland; and the papers cited therein.
\bibitem{[2]} See e.g. J. Bulmahn, Talk given at the 
above-mentioned conference and the papers cited therein. 
\bibitem{[3]} See e.g.  M.~Arneodo et al. Nuc.Phys. {\bf B429} (1994) 503. 
\bibitem{[4]} See e.g.   M.R.~Adams et al. 
Z. Phys. {\bf C65}, 225 (1995). 
\bibitem{[5]} See e.g. the talks given at the   International Workshop on Deep
Inelastic Scattering and Related Phenomena, April 1996, Roma, Italy; and the
papers cited therein.
\bibitem{[6]}  F.E.~Low, Phys. Rev. {\bf D12}, 163 (1975);
S.~Nussinov, Phys. Rev. Lett. {\bf 34}, 1286 (1975) and
Phys. Rev. {\bf D14}, 246 (1976); C. Boros, Z. Liang and T. Meng,
Phys. Rev. {\bf D54}, 6658(1996).
\bibitem{[7]}   S.~Aid et al.  Phys. Lett. {\bf B354},
494 (1995), M.~Derrick et al.  Phys. Lett. 
{\bf B345}, 576 (1995); K.Prytz, Phys.Lett. {\bf B311}, 286 (1993) 
and the papers cited therein. 
\bibitem{[8]}  See e.g. K.~Gottfried and V.~F.~ Weisskopf: Concepts of
Particle Physics,
   Vol. II,  Oxford University Press, New York (1986) p.247.  
\bibitem{[9]} P.~M\"attig, Phys. Rep. {\bf 177}, 142 (1989). 
\bibitem{[10]} B. Anderson et al., Phys. Rep. {\bf 97}, 31 (1983),  
T. Sj\"otrand, Comp. Phys. Comm. {\bf 39}, 347 (1986).
\bibitem{[11]}  T.~Henkes (R608 Collaboration), Phys. Lett. {\bf B 283}, 
155 (1992).
\bibitem{[12]} V.~Kubarovsky, Talk given at same conference as in Ref.1 
and 2;  
and  private communications.
\bibitem{[13]} See, e.g., K. Heller,
in {\it High Energy Spin Physics,} Proc. of the 9th Inter. Symp., Bonn,
Germany, 1990, edited by K.~H.~Althoff, W.~Meyer, p.97 Springer-Verlag, 
(1991); and Talk given at the 12th Int. Symp. on High Enrgy Spin Physics, 
Amsterdam (1996). 
\bibitem{[14]}  C.~Boros, Z.~Liang and T.~Meng,
             Phys. Rev. Lett. {\bf 70}, 1751 (1993),
              Phys. Rev. {\bf D51},4867 (1995) and the papers 
              cited therein.
\bibitem{[15]} J.~Felix et al. Phys.~Rev.~Lett.~{\bf 76}, 22 (1996).
\bibitem{[16]} G.~Gustafson and J.~H\"akkinen, Phys. Lett. {\bf B 303} 
(1993) 350. 
\bibitem{[17]} ALEPH-Collaboration; D.~Buskulic et al., Physics Letters 
{\bf B 374} (1996) 319. 
\bibitem{[18]} ZEUS-Collaboration; 
M.Derrick et al. Physics Letters {\bf B 315} (1993) 481-493 

\end{thebibliography}
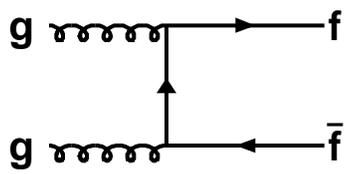
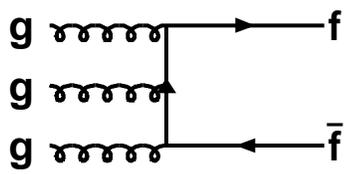
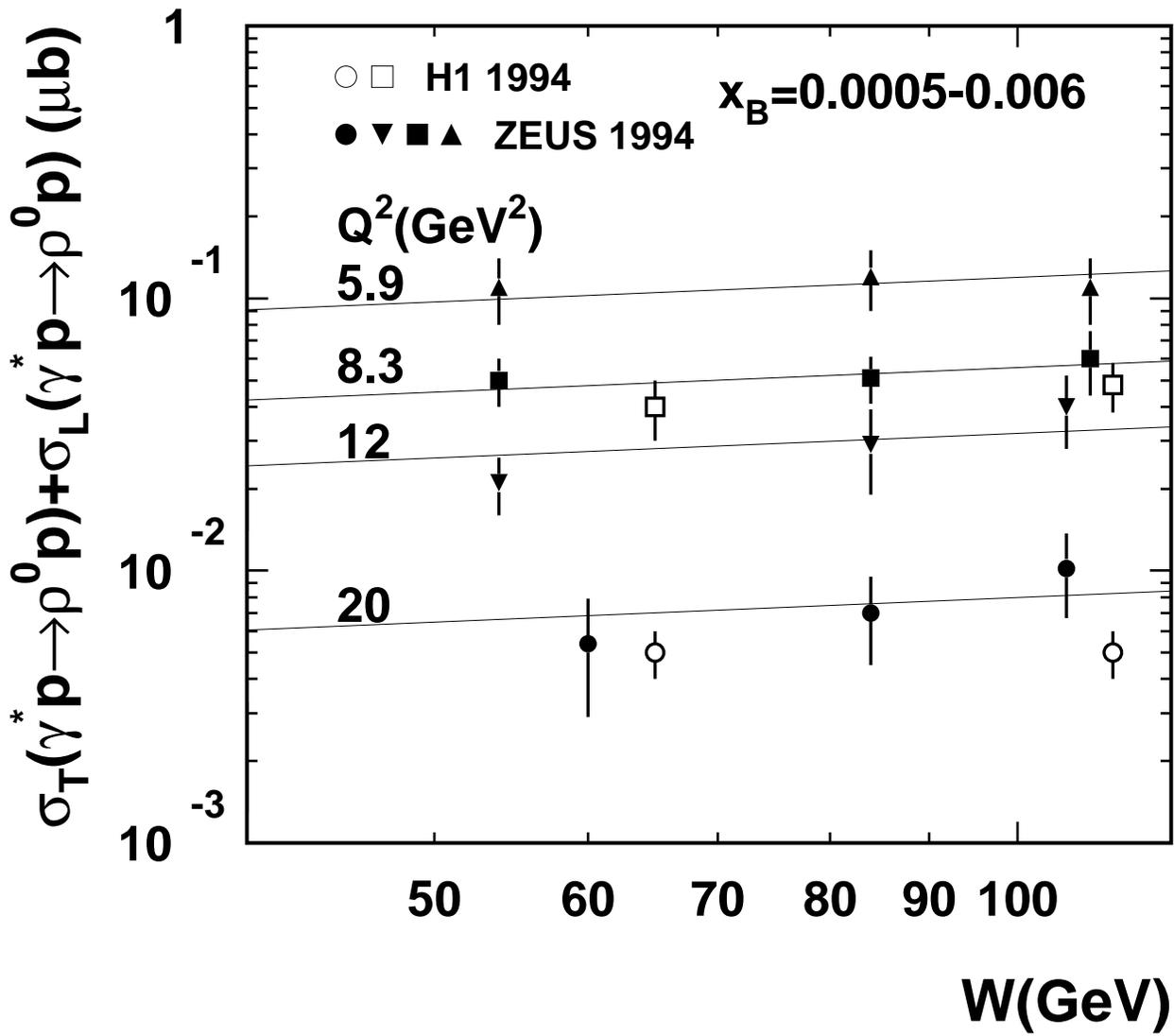
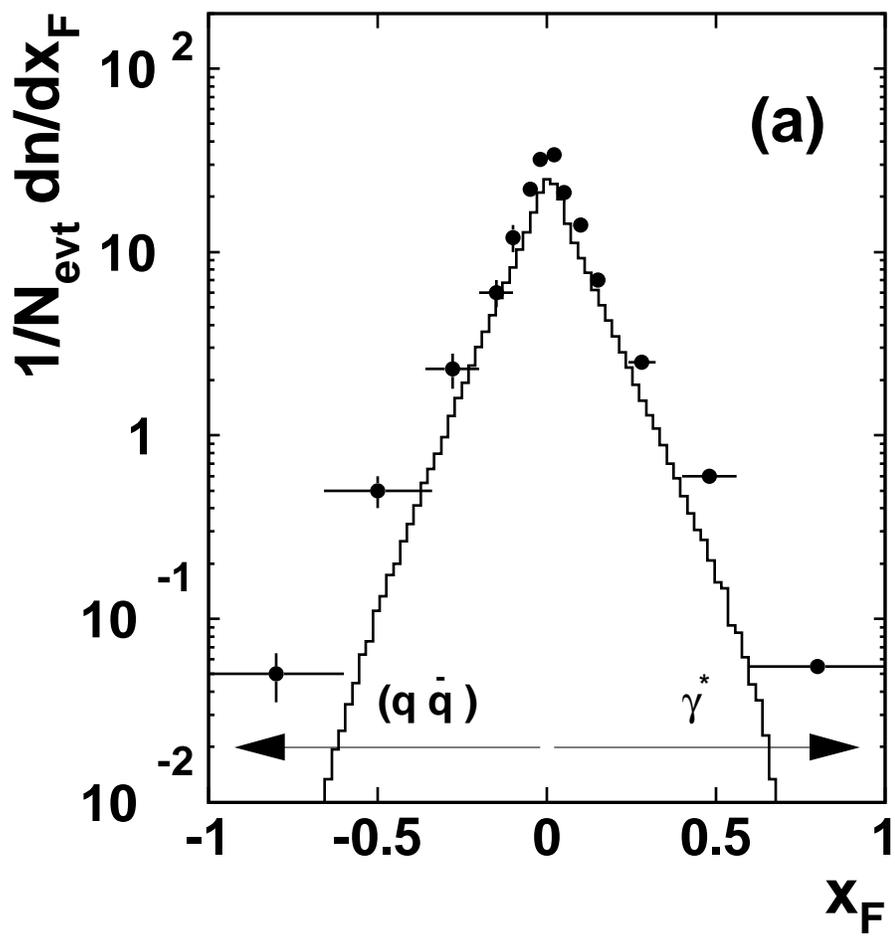
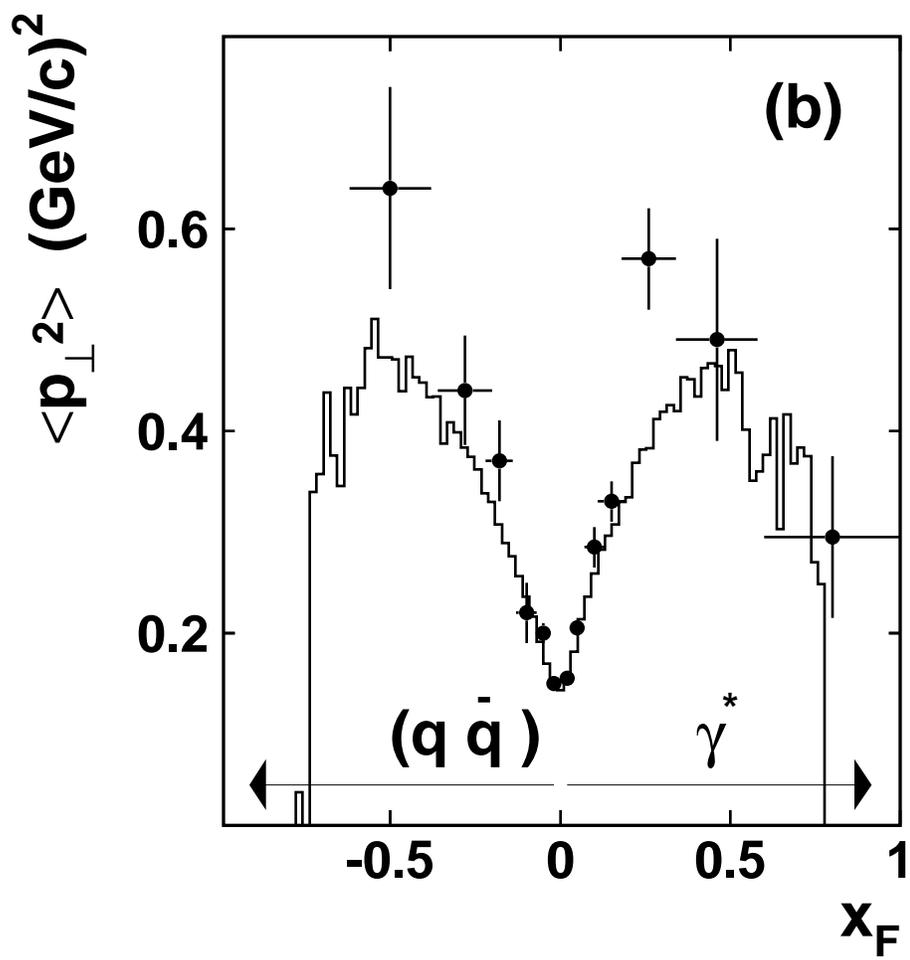
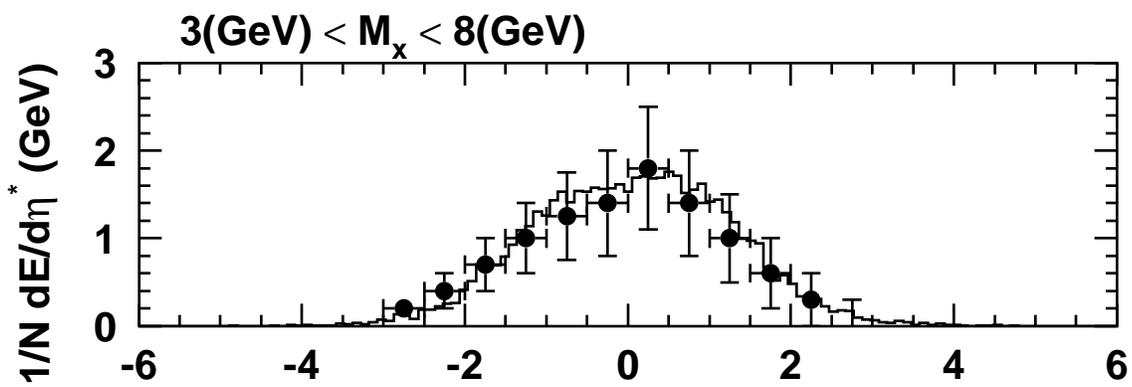
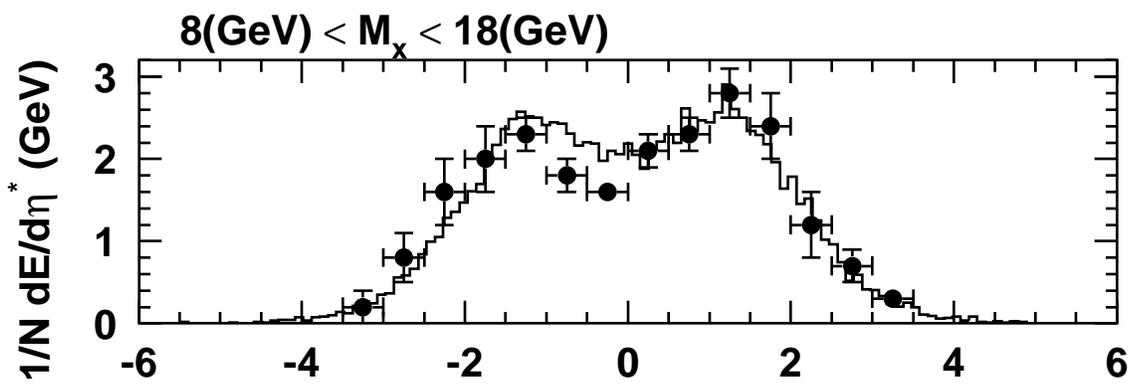
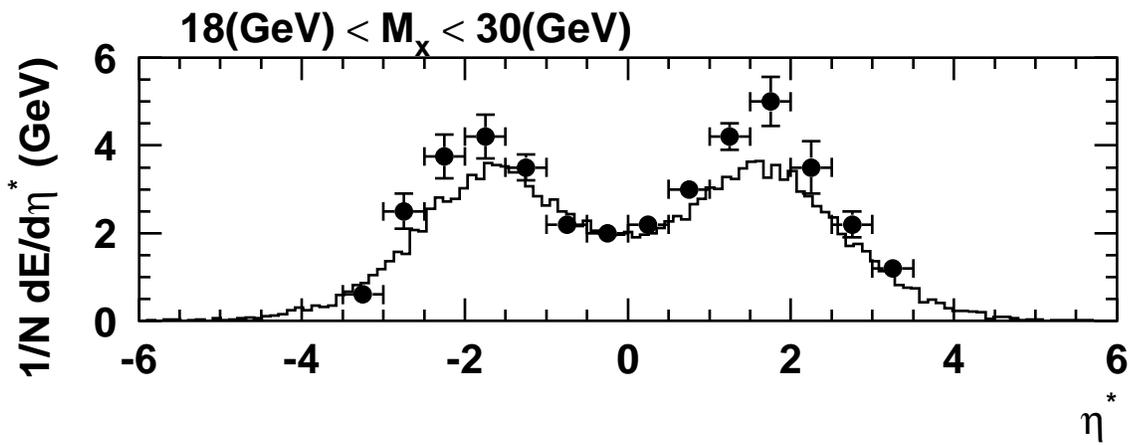
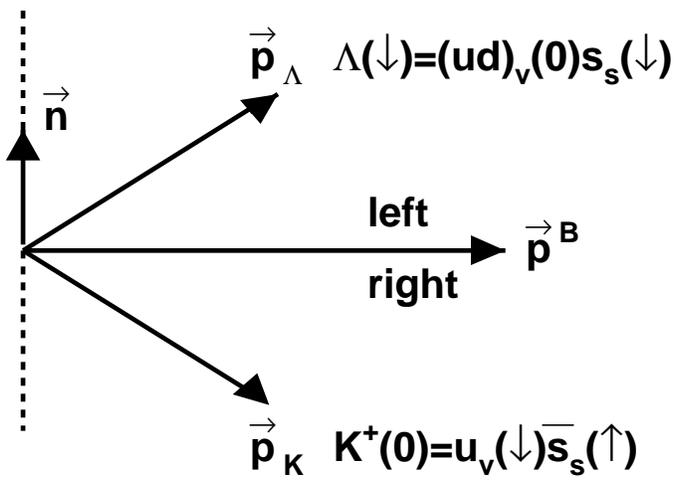
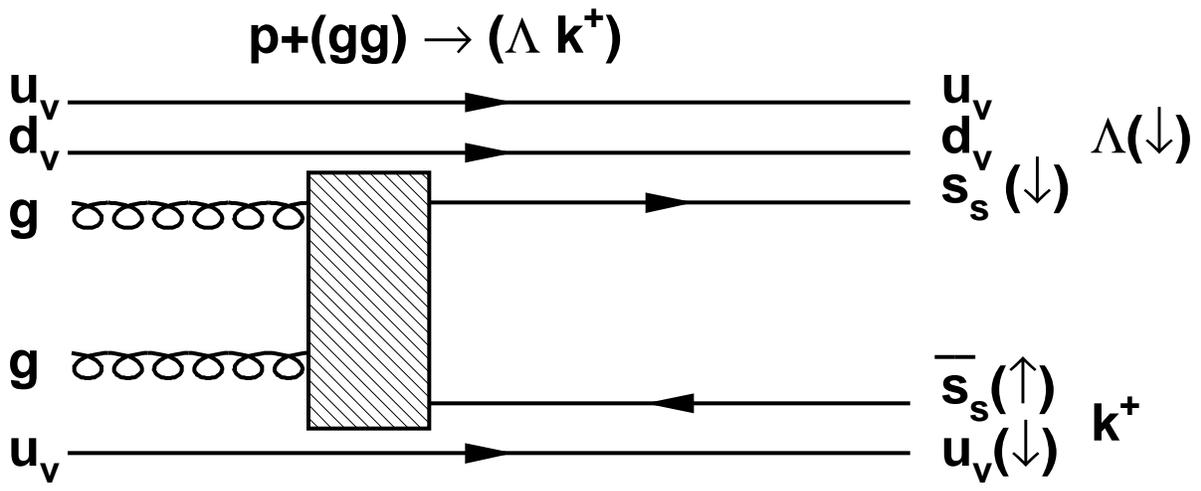
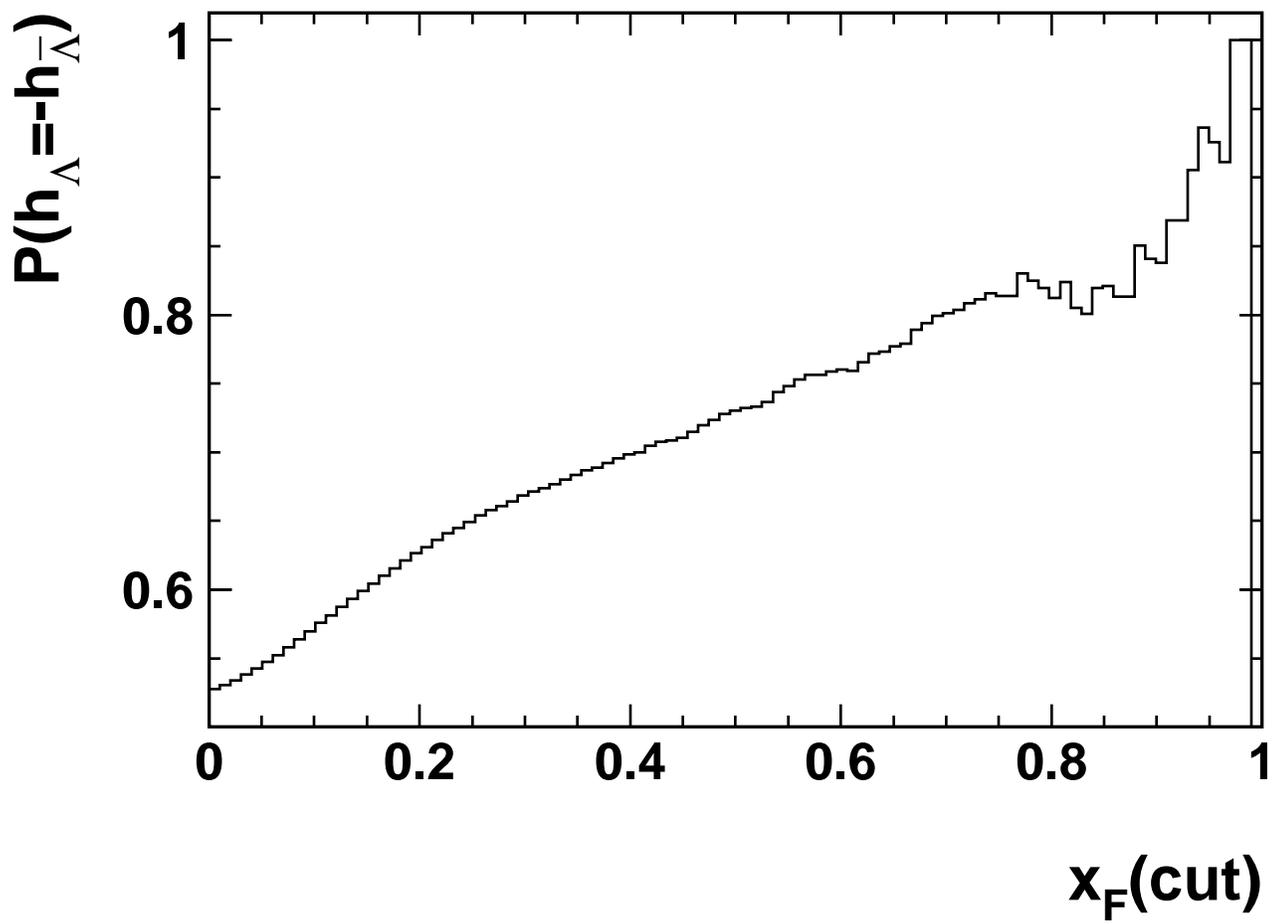
\end{document}